\documentclass[12pt]{article} 
\usepackage{graphicx}
\title{The successive formation and disappearance of density structures in simple expanding systems}
\author{A.R.~Karimov\\
\thanks{E-mail:alexanderkarimov999@gmail.com}
Institute for High Temperatures, \\
Russian Academy of Sciences,\\
Izhorskaya 13/19, Moscow 127412, Russia \\
and Department of Electrophysical Facilities, \\ National Research Nuclear University MEPhI, \\
Kashirskoye shosse 31, Moscow, 115409, Russia
\\ H.~Schamel\\
Physikalisches Institut, Universitaet Bayreuth, \\
D-95440 Bayreuth, Germany}
\date{ }
\begin{document} 
\maketitle
\newcounter{graf}
\begin{abstract}
The Lagrangian fluid description is employed to solve the initial value problem for one-dimensional, compressible fluid flows represented by the Euler-Poisson system. Exact nonlinear and time-dependent solutions are obtained, which exhibit a variety of transient phenomena such as a density collapse in finite space-time or the appearance and, for the first time, a successive dissolution of wavelet structures during the same event. The latter are superimposed on the gross density pattern in the course of the uni-directional expansion of an initially localized density hump. Whereas self-gravitating fluids will always experience collapse, neutral fluids and fluids with repulsive forces, such as a non-neutral, pure electron fluid, can exhibit an evolution of the second type, being determined by the initial conditions. For an electron fluid, being embedded in a neutralizing ion background, these nonlinearities are, however, strongly diminished due to the omnipresent plasma oscillations, which weaken and alternatively change the sign of the collective force.\\
\end{abstract}
\maketitle

\section{Introduction}

There are a lot of many-body systems of different physical nature where the clustering of matter from smooth or almost smooth distributions of particles can occur (see, for example, \cite{ber} - \cite{lk}). Such a process typically proceeds in  particle ensembles with long-range forces, such as gravity or electric interactions, when the clustering leads to the formation of stable spatially inhomogeneous structures \cite{zel} - \cite{dn}. These processes define the system's dynamics and existence. For example, the pancake and adhesion models predict the persistence of a density peak at the flow caustics which appears from the initial distributions having Gaussian initial fluctuations and is due to the development of the gravitational instability \cite{zel,arn}. In this case the singular behavior makes it impossible to reverse the process through the destruction of such structures. One may say that singularities are the factor which stabilizes the space structures. The formation of singularities is a necessary condition for the implementation of these clustering mechanism (see, for example, Refs. \cite{arn} - \cite{yt}). 

It should be noted that a similar behavior can be expected for ideal and/or dissipative plasmas. An example is the generation of a fast ion peak in the plasma expansion problem which is caused by the density collapse of the ion fluid for negligible dissipation becoming a stabilized fast ion peak at the front edge of the expanding plasma, when dissipation is included \cite{ss85,ss}.

Considering the analogy between gravitational and electrostatic fields, it may seem that such nonlinear dynamics is realized in the case of the Coulomb interaction too. However, in the Coulomb systems or the force-free systems there exists another possibility for structure formation without a singular (blow-up) behavior. The dynamics for some initial conditions can lead to the formation of dynamical space structures \cite{k,ks} in which the initial conditions play the role of a driving parameter precluding the formation of singularities but providing the formation of transient dynamical structures. 

The present paper considers some aspects of such a transient clustering mechanism in one-dimensional geometry which, of course, looks like somewhat artificial since all real systems, as a rule, are not one-dimensional ones, in higher spatial dimensions, a large variety of nonlinear dynamical structures can be expected (see, for example, \cite{nat,s04}). Nonetheless, this model indicates what may come about in real systems. In the case worked out there are no stabilizing factors such as dissipative processes or formation of caustics. So one can expect that the structure emerged may break down or simply disappear. It means that some smooth initial density distributions will produce a wave-like structure which over time will emerge and subsequently disappear yielding a distribution becoming close to the initial distribution. We shall consider such a possibility and we try to define the conditions for which such an evolution of the one-dimensional system may be realized. 

\section{Model}

Here, we consider two types of one-dimensional, time-dependent patterns of infinite, compressible medium. First type is the medium containing only movable, particles which interact with each other via self-consistent gravitational (or Coulomb) fields. Another type of the medium is a cold plasma containing movable electrons and immobile ions of constant density $n_i\equiv 1$. In both media, the structure formation comes about because of the self-consistent forces caused by the distribution of the movable particles within the system and the effect of initial conditions. 

The dynamics of these systems is governed by the Euler-Poisson hydrodynamic equations written in dimensionless variables as
\begin{equation} 
{\partial n \over \partial t} + 
{\partial \over \partial x}(n U)=0\/, 
\label{1_nature} 
\end{equation} 
\begin{equation} 
{\partial U \over \partial t} + U 
{\partial U \over \partial x}= 
\delta {\partial \Phi\over \partial x}\/, 
\label{2_nature} 
\end{equation} 
\begin{equation} 
{\partial^2 \Phi\over \partial x^2}= (n - \epsilon)\/, 
\label{3_nature} 
\end{equation} 
where $n$ is the density of movable component and $U$ the velocity of the fluid. The self-consistent potential $\Phi$ corresponds for constant $\delta=-1$ to an attractive force and for $\delta=+1$ to a repulsive force in the system, whereas the case $\delta=0$ describes an ordinary flow without potential forces. In our model the constant $\epsilon$ can take two values: $\epsilon=0$ for the case of zero background describing one component flow with pure gravitational or Coulomb interaction and $\epsilon=1$ for a nonzero background describing the evolution of cold plasma medium. 

It is also worth noting that though the Euler-Poisson system appears in numerous applications including plasma physics and astrophysical applications there is a sense to consider it as an inviscid limit of Burges turbulence model \cite{bur} in the systems with collective interaction. From such point of view the influence of initial conditions on the regular and chaotic structures formation has been of great interest. 

In order to study this effect we consider an initial-value problem for the equations (\ref{1_nature})--(\ref{3_nature}) in the infinite interval $-\infty \leq x \leq \infty$. (This allows us to avoid the influence of boundary conditions and finite size of the interval on the dynamics.) Presuming that the initial velocity profile of movable particles is specified by the dependence 
\begin{equation} 
U(x,t=0) =  U_D(x) + B\sin(kx) \equiv U_0(x), 
\label{4_nature} 
\end{equation}
where $B$ and $k$ are the arbitrary positive constants. 
$U_D(x)$ will be smooth and monotonic  and we assume that the sign of the velocity derivative,  $U_0^{\prime}(x)$, is retained. Second term in (\ref{4_nature}) describes some regular perturbation in the initial conditions. The initial density profile of the movable component is assumed to be an even function 
\begin{equation} 
n(x,t=0) = n_0(x)\/,  
\label{5_nature} 
\end{equation}
which preserves a finite mass in the interval $-\infty \leq x \leq \infty$, i.e. 
$$\int_{-\infty}^{+\infty} n_0(x)dx < \infty\/.$$
Moreover, we must supplement the above-given relations by the following conditions, arising from the symmetry of the
problem at $x=0$:
\begin{equation} 
\left. {\partial \Phi\over \partial x}\right|_{x=0} = 0, \hspace{11mm} U( x=0,t)=0\/. 
\label{symmetry_nature} 
\end{equation}

Now we transform the original system (\ref{1_nature})--(\ref{3_nature}) to a simpler form. Using the definition
\begin{equation} 
N(x,t)=\int_0^x n(x^{\prime},t) dx^{\prime}\/  
\label{4_col}
\end{equation}
and condition (\ref{symmetry_nature}), we can conveniently cast (\ref{1_nature})--(\ref{3_nature}) into the form 
\begin{equation} 
{\partial N \over \partial t} + U {\partial N \over 
\partial x}=0\/,
\label{6_col}
\end{equation}
\begin{equation} 
{\partial U \over \partial t} + U {\partial U 
\over \partial x} = \delta (N -\epsilon x)\/.
\label{7_col}
\end{equation}

\section{Initial-value problem in the Lagrangian frame}

The study of the system (\ref{6_col})--(\ref{7_col}) is conveniently carried out in the Lagrangian variables (see, for example, \cite{dav}):
\begin{equation} 
\tau=t,\hspace{11mm} \xi=x-\int_0^t U(\xi, t^{\prime})dt^{\prime}\/,
\label{1_sh}
\end{equation}
where $x(\xi, t)$ satisfies the initial condition
$$x( \xi, 0)=\xi$$
and provides
\begin{equation} 
U(\xi, \tau) =\left({\partial x \over \partial 
\tau} \right)_{\xi}\/.
\label{U_nature} 
\end{equation}
This transformation holds true when the Jacobian (see, for example, \cite {s04})
$$J( \xi,\tau) \equiv \left({\partial x( \xi, \tau) \over \partial 
\xi} \right)_{\tau}\/ = 1 +\int_0^\tau U^{\prime}(\xi, t^{\prime})dt^{\prime}
  > 0\/,$$
besides, the condition of sign conservation of  $J( \xi,\tau)$ 
eliminates singularities of the flow.\\ 

Under the transformation (\ref{1_sh}) the Eqs. (\ref{6_col}) and (\ref{7_col}) with (\ref{4_col}) in hand turn into 
\begin{equation} 
{\partial \over \partial \tau} N( \xi,\tau) = 0\/,
\label{9_col}
\end{equation}
\begin{equation} 
{\partial U \over \partial \tau} = \delta (N - \epsilon x)\/.
\label{10_col}
\end{equation}
Eq.(\ref{9_col}), first of all, proves that  $N( \xi,\tau)$ is an invariant of the flow, which also holds true for $U( \xi,\tau)$ when $\delta =0$.

To simplify (\ref{9_col})--(\ref{10_col}) further we derive some additional relations being connected with the basic parameters of the problem. Differentiating the relation (\ref{4_col}) with respect to $\xi$, switching thereby to the mass variable $N$ as a Lagrangian variable  instead of $\xi$ (see, for example, \cite {ss,s04}), we get
\begin{equation} 
{\partial x \over \partial N} = V\/,
\label{5_col}
\end{equation}
where we define the quantity $V=1/n$ as the specific volume (see, for example, \cite{ss}). Finally, after differentiating (\ref{5_col}) with respect to $\tau$ and bearing in mind (\ref{U_nature}) we find
\begin{equation} 
{\partial U \over \partial N} = {\partial V \over 
\partial \tau}\/.
\label{12_col}
\end{equation}
Relations (\ref{10_col}) and (\ref{12_col}) can be used to obtain a wave equation for the specific volume $V=V(N,\tau)$ instead of the system (\ref{1_nature})--(\ref{3_nature}). Differentiating the relation (\ref{10_col}) with 
respect to $N$ and taking into account (\ref{12_col}), we obtain 
\begin{equation} 
{\partial^2 V \over \partial \tau^2} = \delta (1 - \epsilon V)\/,
\label{13_col}
\end{equation}
which should be supplemented by the initial conditions:
\begin{equation} 
V(N,\tau=0) = V_0=1/n_0\/, 
\label{V0}
\end{equation}
\begin{equation} 
\left.{\partial V \over \partial \tau}\right|_{\tau=0} = 
\left.{\partial U \over \partial N}\right|_{\tau=0} =
\left. \left[{\partial N \over \partial 
\xi}\right]^{-1}{\partial U \over \partial 
\xi}\right|_{\tau=0} = V_0{\partial U_0 \over \partial \xi}\/.
\label{dV0}
\end{equation}

\section{General solutions and their applications}

Depending on the parameter $\epsilon$, the equation (\ref{13_col}) has got different solutions which describe different physical situations. Indeed, in the case $\epsilon=0$ when in the system there is only a gravitational or electrostatic interaction, the general solution of the problem (\ref{13_col})-(\ref{dV0}) is then finally provided by
\begin{equation} 
n(\xi,\tau)=n_0(\xi) \left[1 + 
{\partial U_0 \over \partial \xi} \tau + \delta n_0(\xi) {\tau^2 \over 2}\right]^{-1}\/ \equiv {n_0(\xi) \over J(\xi,\tau)}.
\label{cold_col}
\end{equation}
We mention that the expression for $J(\xi,\tau)$ can also be directly  obtained from the solutions of Eqs. (\ref{9_col}) and (\ref{10_col}), namely $N(\xi,\tau)=N_0(\xi)$ and $U(\xi,\tau)=U_0(\xi) + \tau \delta N_0(\xi)$, and by utilizing the definitions of $N$ and $J$. The basic solution of our problem in the $(\xi,\tau)$- representation is hence given by (\ref{cold_col}) and
\begin{equation} 
x(\xi, \tau) =\xi + \tau U_0(\xi)  + \frac{\tau^2}{2} \delta \int_0^\xi n_0(\xi^{\prime})d\xi^{\prime} \/,
\label{x_sh}
\end{equation}
\begin{equation} 
U(\xi, \tau) =U_0(\xi) + \tau \delta \int_0^\xi d\xi^{\prime}  n_0(\xi^{\prime})\/.
\label{U_sh}
\end{equation}
That is the well-known solutions which have been more than once used to study the some astrophysical problems (see, for example, Refs. \cite{zel,arn,ber}). The non-neutral plasma systems such as different the charged particle traps \cite{dn,dn_1,anderegg}, or the charged particle beams \cite{andr}-\cite{rsr}, or the electron transport in planar diodes \cite{s04} are natural media which are described by these relations for $\delta=1$.

As mentioned, for $\delta=-1$ such type of solutions geometry describes the formation of finite-time singularities owing to the development gravitational instability. In three-dimensional geometry these singularities form cellular structures being concentrated along the cell boundaries \cite{zel}. For $\delta=1$ it possible to escape the development of singular behavior and here we focus on such issue. 

In the case $\epsilon=1$ and $\delta=1$ one can have
\begin{equation} 
n(\xi,\tau)= n_0(\xi) \left[n_0(\xi) + {\Bigl (}1 - n_0(\xi) {\Bigr )} \cos(\tau) + {\partial U_0 \over \partial \xi} \sin(\tau)\right]^{-1}\/.
\label{d_dav}
\end{equation}
The relation (\ref{5_col}) can be rewritten as 
$${\partial x \over \partial N}=
{\partial x \over \partial \xi}\left[{\partial N \over \partial \xi}\right]^{-1}
={1 \over n_0}{\partial x \over \partial \xi}=V\/.$$
Then from this relation and (\ref{d_dav}) we find
\begin{equation} 
x(\xi, \tau) = \xi + {\Bigl (}1 - \cos(\tau) {\Bigr )} \left[\int_0^\xi n_0(\xi^{\prime})d\xi^{\prime} - \xi \right] + U_0(\xi)\sin(\tau)\/.
\label{x_dav}
\end{equation}
from which and (\ref{U_nature}) it follows
\begin{equation} 
U(\xi, \tau) = U_0(\xi)\cos(\tau) + \left[\int_0^\xi n_0(\xi^{\prime})d\xi^{\prime} - \xi\right]\sin(\tau)\/.
\label{Udav}
\end{equation}

The solutions (\ref{d_dav})-(\ref{Udav}) have been studied in detail by Dawson \cite{daw}, Kalman \cite{klm}, Davidson and Schram \cite{ds,dav} to describe the dynamical features of nonlinear Langmuir oscillations in the different limits of a cold plasma. These findings should be helpful in identifying or interpreting cold-plasma expansion phenomena found in space and laboratory experiments \cite{ler} - \cite{wo}. In works \cite{ds,dav}, the special case of initial velocity $U_0^{\prime} \equiv 0$ and the electron density $n_0=1+\Delta\cos(k\xi)$ (in our notation) was considered, the similar issue was studied for a periodic, step-like initial density distribution in \cite{klm}. It has been shown that this distribution $n_0$ propels the periodic wave structure of density in extremum points $x= i\pi/k$, where $i=0, \pm 1, \pm 2, \ldots$. In the framework of initial-value problem, the initial velocity perturbations of special kind ($\sim \sin(k\xi)$) in spatially homogeneous plasma ($n_0 \equiv 1$) have been discussed by Kalman \cite{klm}. As was established, this process can lead to the formation of standing density waves whose form is determined by the form of the initial velocity profile. 

Now based on the relations (\ref{cold_col})-(\ref{U_sh}) and (\ref{d_dav})-(\ref{Udav}), we show that for $U_0^{\prime} \ne 0$ and $n_0 \ne 1$ it is possible the formation of density structures differing from the above-described variants. Moreover, we shall show that the structure formation in the cases $\epsilon=0$ and $\epsilon=1$ occurs in different way.

According to \cite{k}, the dependence of the density on the coordinate $x$ for any moment of time $t$ is qualitatively similar to the dependence $n(\xi, \tau)$ in Lagrangian coordinates on $\xi$ at any moment of time $t=\tau$. Therefore, all the information on the characteristics of the density distribution is contained in (\ref{cold_col}) and (\ref{d_dav})  or in theirs derivatives $ \partial n / \partial \xi $. Expression (\ref{cold_col}) and (\ref{d_dav}) can then be used to obtain 
\begin{equation} 
{\partial n \over \partial \xi} =g_{\epsilon}(\tau) { n_0^\prime(\xi) \over J^2(\xi, \tau)}
\left[h_{\epsilon}(\tau) - \theta_0(\xi)\right]\/,
\label{d_nature}
\end{equation} 
where 
\begin{equation} 
\theta_0( \xi)= {n_0 \over n_0^{\prime}}U_0^{\prime \prime} - U_0^{\prime}
\label{f_nature}
\end{equation}
and functions $g_{\epsilon}(\tau)$ and $h_{\epsilon}(\tau)$ are defined as 
\[\begin{array}{llcl}
g_{\epsilon}=\tau,      & h_{\epsilon}=1/\tau,      &{\rm for} & \epsilon=0, \\
g_{\epsilon}=\sin(\tau),& h_{\epsilon}=1/{\rm tg}(\tau) &{\rm for}&  \epsilon=1.
\end{array}\]
It is easy to see that $g_1 \to g_0$ and $h_1 \to h_0$ under $\tau \to 0$. 

As is seen from (\ref{d_nature}) for $J \ne 0$ and $ g_{\epsilon} \ne 0$, the function $n(\xi,\tau)$ reaches the local extremum with respect to $\xi$ at time $t = \tau$ in points where $n_0^{\prime}=0$ and points $\xi_s$ being the roots of the algebraic equation
\begin{equation} 
\theta_0(\xi) = h_{\epsilon}(\tau)\/,
\label{7_nature}
\end{equation}
where $\tau$ must be treated as a parameter. As is seen, the number of points of such type depends on the form of functions $\theta_0(\xi)$ and $h_{\epsilon}(\tau)$ and this number can be varied with respect to time. This suggests that under some circumstances we can get a wavelet structure from the initially monotonic density profile and then the one can be destroyed when the Eq. (\ref{7_nature}) has no solution.

\section{The structure dynamics for $\epsilon=0$}

We start from case $\epsilon=0$ which refers to the systems with a pure electrostatic or gravitational interactions. The case of the force-free systems must be considered as a limit of these systems.

In order reveal the structure of $\theta_0(\xi)$ we substitute (\ref{4_nature})  and (\ref{5_nature})  into (\ref{f_nature}) and rewrite this result in the following form 
$$\theta_0(\xi) = G(\xi) +\nu(\xi)\/,$$
where 
\begin{equation} 
G(\xi)= {n_0 \over n_0^{\prime}}U_D^{\prime \prime} - U_D^{\prime}
\label{G_nature}
\end{equation}
and
\begin{equation} 
\nu( \xi)= -kB\left[{n_0 \over n_0^{\prime}}k\sin(k\xi) + \cos(k\xi)\right]\/.
\label{nu_nature}
\end{equation}
It is easy to see from these relations that Eq. (\ref{7_nature}) can have a lot of roots for monotonic initial distributions $n_0(\xi)$ and $U_0(\xi)$ when the second derivative of the velocity is non monotonic. In this case the form of $\theta_0(\xi)$ may be essentially defined by the perturbative part of the velocity via function $\nu(\xi)$ having a wavelet structure, i.e. it consists of local minima and maxima. Noting that $\nu(\xi)$ reaches its local extrema in the points $\xi_s$ being the roots of equation $ d \nu(\xi)/ d \xi =0$, e.i.
$${\rm tg}(k\xi_s)= k {n_0^{\prime} \over n_0^{\prime\prime}}\/.$$
However, the manifestation of $\nu(\xi)$ in $\theta_0$ completely depends on the magnitude and sign of the function $G(\xi)$. Thus, the possible way of evolution for the system is determined by the relation between the functions $G(\xi)$ and $\nu(\xi)$. 

In \cite{ks,kss} we applied such approach to study the formation of wavelet structures under the influence of the perturbative part of the initial velocity distribution when the function $\theta_0(\xi)$ can have both negative and positive values. Here we consider the restricted case when the governing function $\theta_0(\xi) > 0$, and discuss the physical consequences of this restriction. This regime is of interest since it admits both clustering and dissolution of the pattern arising from smooth initial distributions. So we proceed to find out the initial conditions when such regimes can exist.

First of all, we shall find out the conditions for the initial data under which the singular behavior has been eliminated. In the present case the Jacobian takes the form:
\begin{equation} 
J(\xi,\tau) = 1 + \left[U_D^{\prime} + kB \cos(k\xi)\right] \tau + {\delta n_0\over 2}\tau^2\/.
\label{1_jacob}
\end{equation}

From the condition $J = 0$ we get the quadratic equation
$${\delta n_0\over 2} \tau^2 + \left[U_D^{\prime} + kB \cos(k\xi)\right] \tau + 1 = 0\/,$$
whose positive solutions define the time $\tau_*$ when the first collapse may appear. The roots of this quadratic equation are determined by
\begin{equation} 
\tau_* = {-\Lambda \pm \sqrt{\Lambda^2 - 2\delta n_0}\over \delta n_0}\/,
\label{qua}
\end{equation}
where 
\begin{equation} 
\Lambda = U_D^{\prime} + kB\cos (k\xi)\/.
\label{lambda}
\end{equation}
As seen from (\ref{qua}) it is impossible to escape from the formation of singularities when we have a self-gravitating fluid, $\delta = -1$. Therefore, we focus our attention on the case $\delta = 1, 0$ where we have no positive roots of quadratic equation when $\Lambda >0$. This condition is always fulfilled for  
\begin{equation} 
U_D^{\prime} > kB\/.
\label{kb}
\end{equation}

In order to study the influence of the initial conditions on $\theta_0(\xi)$ now we have to define the form of $n_0(\xi)$  and $U_D(\xi)$ which make it possible to avoid a singular dynamics and provide a wavelet structure for positive $\theta_0(\xi)$. As is seen from (\ref{nu_nature}) in the case
\begin{equation} 
\left| {n_0 \over n_0^{\prime}}\right| \leq M\/,
\label{1_n0a}
\end{equation}
where $M$ is some positive constant, it holds 
\begin{equation} 
\mid \nu(\xi) \mid \leq \nu_{m}:= kB (1 + kM)
\label{2_n0a}
\end{equation}
for any $\xi$. So, taking into account the definitions (\ref{G_nature}) and (\ref{nu_nature}) we conclude that the condition 
$$\theta_0(\xi) >0$$ 
is satisfied provided that there exists  such a function $U_D(\xi)$ for which the inequality 
\begin{equation} 
{n_0 \over n_0^{\prime}} U_D^{\prime\prime} - 
U_D^{\prime} \geq \nu_m
\label{3_n0a}
\end{equation}
holds for any $\xi$. 
Here we restrict ourselves to the simplest case when the inequality (\ref{3_n0a}) turns into equation 
\begin{equation} 
{n_0 \over n_0^{\prime}} U_D^{\prime\prime} - 
U_D^{\prime} = \nu_m
\label{4_n0a}
\end{equation}
which has  solution
\begin{equation} 
U_D(\xi) = C + \int_0^{\xi} n_0(x) dx - \nu_m\xi\/,
\label{5_n0a}
\end{equation}
where $C$ is an arbitrary constant. This relation determines an admissible type of $U_D$ which is defined by initial profile $n_0$.

Inserting (\ref{5_n0a}) into (\ref{kb}), we obtain the relation
\begin{equation} 
n_0(\xi) \geq kB (2 +kM)
\label{6_n0a}
\end{equation}
which together with Eq.(\ref{3_n0a}) impose a restriction on the possible form of initial distribution $n_0$. Combining these expressions we get 
\begin{equation} 
kB (2 +kM) \leq \left| {n_0 \over n_0^{\prime}} \right| \leq M\/.
\label{7_n0a}
\end{equation}
For $ n_0^{\prime} < 0$ having physical meaning the condition (\ref{7_n0a}) is reduced to
$$ - {1 \over kB (2 +kM)} \leq {n_0^{\prime} \over n_0} \leq -{1 \over M}\/.$$
Integrating this relation we obtain
\begin{equation} 
\exp\left[- {\xi \over kB (2 +kM)}\right] \leq n_0(\xi) \leq \exp\left[- {\xi \over M}\right]\/.
\label{8_n0a}
\end{equation}
It is also worth emphasizing that this relation has a meaning only for 
\begin{equation} 
0 < kB (2 +kM)  <  M\/.
\label{9_n0a}
\end{equation}
Thus, the relations (\ref{5_n0a}), (\ref{8_n0a}) and (\ref{9_n0a}) define some non-empty set of the governing parameters  $k$, $B$ and $M$ when nonsingular dynamics occurs and the positive governing function $\theta_0(\xi)$ has a wavelet form. An explicit example in the 3D parameter space $(k,B,M)$ for which (\ref{9_n0a}) is valid is given by $(0.5, 1, M>4/3)$.

This leads us to the principal features in the dynamics in contrast to \cite{ks,kss} where $\theta_0(\xi)$ can change its sign. For definiteness, we display such dynamics with the initial profile of form
\begin{equation} 
n_0(\xi) = \exp(- \mid\xi\mid/L)\/,
\label{10_n0a}
\end{equation}
where constant $L$ satisfies all necessary conditions, noting that for an initial Gaussian density profile such a process of wave let creation and dissolution would be absent.

\begin{figure} 
\begin{center}
\includegraphics[width=11.cm]{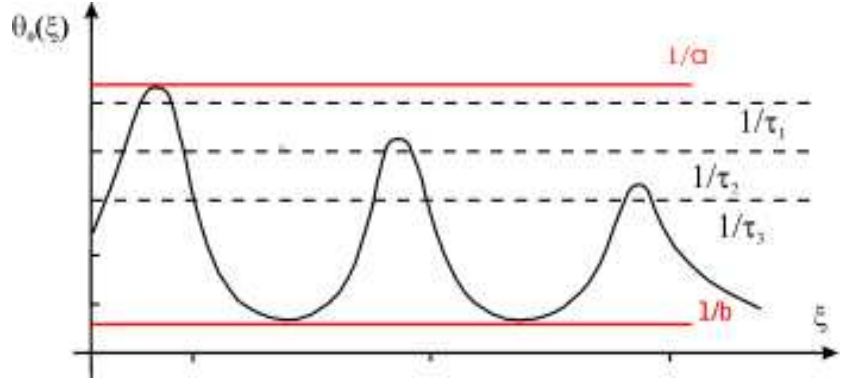}  
\end{center}   
\caption{The qualitative solution of Eq. (\ref{7_nature}) for $\theta_0(\xi) >0$ \label{fig_1}}
\end{figure}

\begin{figure} 
\begin{center}
\includegraphics[width=11.cm]{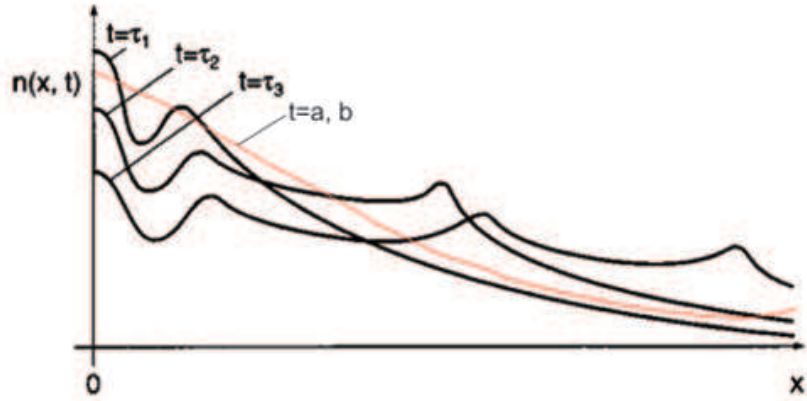}  
\end{center}   
\caption{The qualitative behavior of the density corresponding to Fig. \ref{fig_1} \label{fig_2}}
\end{figure}

Fig. \ref{fig_1} depicts the qualitative dependence of $\theta_0(\xi)$ which may be typical for profile of  type  (\ref{10_n0a}). In this situation the formation of wavelet structure on the density profile in the course of time occurs similar to the case studied in \cite{ks} for small times when the initial smooth and monotonic distribution (\ref{10_n0a}) is transformed into density profile with a large number of local extrema as  sketched in Fig. \ref{fig_2} for $\tau > a$. Such script scenario will be implemented until the Eq. (\ref{7_nature}) will have a real root. However, it is clear that owing to $\theta_0(\xi) >0$ such behavior cannot last indefinitely. There is a moment of time, say $\tau=\tau_{min}$ when we lose one root of the Eq. (\ref{7_nature}), and consequently one hump in the dependence of density profile. It means that the wavelet structure disappears in the course of time. For this $\theta_0$ there always exists a moment of time, say $\tau > b$ when the Eq. (\ref{7_nature}) has no real root anymore and the wavelet structure of density dependence is transformed into a distribution which is similar to the initial distribution. Thus, for the given initial conditions the wavelet structure can exist only for some finite time. The physical reason for this unexpected behavior may be found in the temporal change of the  strength of the nonlinearity, which gives rise to a steepening and structure formation in the earlier stage when it is strong, but later, when it becomes weakened by the expansion process, it  is no longer in the position to sustain the structure, which then dissolves.

In the case presented in Figs. \ref{fig_1} and \ref{fig_2} such a structure exists only in the interval $a \leq t \leq b$. For times $t > b$ the process may be again repeated in general. However, it is impossible to replicate this scenario identically since the new velocity distribution will differ from the initial distribution $U_0(x)$. This property distinguishes the cases $\delta = 1, 0$ from the case $\delta = -1$ in which the formation of singularities eliminates such behavior. In this connection, let us dwell on the time reversibility of the system (\ref{1_nature})--(\ref{3_nature}). It is easy to see that the Eqs. (\ref{1_nature})--(\ref{3_nature}) remain invariant under the transformation
\begin{equation} 
t \to \alpha t, \hspace{7mm} x \to \beta x, \hspace{7mm} n \to {1 \over \alpha^2} n, \hspace{7mm} U \to {\beta \over \alpha} U, \hspace{7mm} \Phi \to {\beta^2 \over \alpha^2} \Phi\/, 
\label{sym_nature} 
\end{equation}
where $\alpha$ and $\beta$ are some parameters of transformation. Now let the time change the direction, i.e. we have to put $\alpha = - 1$ in these relations. As is seen from (\ref{sym_nature}) this leads only to $U \to - U$. It means that (\ref{1_nature})--(\ref{3_nature}) is a reversible system and generally we can expect to get the initial space distribution of density and velocity; at least such a possibility is not prohibited. However, as was shown above, in the present case the destruction of space structure is caused by other reasons. Though the same result follows directly from Eq.(\ref{13_col}), here we use the Euler frame as physically more intuitive one. 

\section{The structure dynamics for $\epsilon=1$}

Now we pass to the plasma structures formation by considering the Vlasov-Poisson system with parameters $\delta=1$ and $\epsilon=1$. Following to \cite{ds} we put $U_0^{\prime} = 0$ then Eq. (\ref{d_nature}) is reduced to
\begin{equation} 
{\partial n \over \partial \xi} = {n_0^{\prime}\cos(\tau)\over [n_0 + (1-n_0)\cos(\tau)]^2}\/.
\label{ds_n}
\end{equation} 
As is seen from this relations, for $\cos(\tau)\ne 0$ and $n_0 + (1-n_0)\cos(\tau) >0$ the number points of exremum coincides with ones of the initial profile $n_0$ as this was established in \cite{ds} and no additional structure may appear in this case.

Assume that $n_0(\xi) \equiv 1$ as this was done in \cite{klm}. From (\ref{d_nature}) it easy to see that in this case there are no conditions for the development of the wave breaking. Then the relation (\ref{d_nature}) is transformed into 
\begin{equation} 
{\partial n \over \partial \xi} = -{ \sin(\tau)\over [1 + U_0^{\prime} \sin(\tau)]^2} U_0^{\prime \prime}\/,
\label{k_n}
\end{equation}
from which it follows that the formation of new extrema is determined by $U_0^{\prime \prime}$. Namely there is an accumulation of particles in the regions where $U_0^{\prime \prime} < 0$ and there is a diminishing density in the regions $U_0^{\prime \prime} > 0$. Also, at the time moments $\tau= \pi i, i=0, 1,  2, \ldots$ we have
$${\partial n \over \partial \xi} = 0\/.$$
This type of exteremum is perfectly caused by the electrostatic electron oscillations of a plasma medium. We now get an oscillatory time behavior with repeated sign reversals of $U(\xi,\tau)$. The temporal evolution is no longer an uni-directional expansion with increasing out-flow velocities, as in the previous cases, but it is now characterized by a time-periodic behavior, where out- and in-flows alternate each other. This is physically expected since in the presence of an ion background the force influencing on the electrons changes periodically its direction.  This is due to a density variation in Euler space as it holds 
$$N(x,t)= N_0(\xi) = N_0{\Bigl (} x - \int_0^t U(x,t') dt'{\Bigr )}\/.$$ 
The frequency of this oscillation is, of course, the plasma frequency, given by $\omega =1$ in our normalization. Thus, at moments $\tau= \pi i, i=0, 1,  2, \ldots$ the spatial structure is destroyed and the initial, uniform density distribution emerges and then the process repeats in some local region of phase space [see Eq. (\ref{x_dav}) and (\ref{Udav})]. 

In general case when $U_0(\xi)\neq 0$ and $n_(\xi) \neq$ const. the formation of density structure comes about by the similar way. The dependence of $n'(\xi, \tau)$ is controlled by the behavior of the function $\theta_0(\xi)$ via Eq. (\ref{d_nature}) with $h_1(\tau)=1/{\rm tg}(\tau)$. In contrast to the case $\epsilon=0$ here we have the periodic, unlimited function of time. This leads to the formation and disappearance of density profile will be repeated with the period of the function $h_1(\tau)$. Since function $h_1(\tau)$ takes both positive and negative values therefore the Eq. (\ref{d_nature}) always has roots for any time $\tau$ for which it holds 
\begin{equation} 
\mid h_1(\tau)\mid \leq {\rm Max}(\mid\theta_0(\xi)\mid) 
\label{time_nature}
\end{equation}
and the formation of density structure can occur only for these times. Outside the time intervals defined by relation (\ref{time_nature}) there is a restoration of the initial profile.

\section{Conclusion}

In this paper we have considered the influence of a initial velocity on the dynamics of one-dimensional Euler-Poisson system. For the Coulomb systems ($\delta=-1$) and the force-free systems ($\delta=0$) it was shown that the wavelet-like density structure can appear and disappear for some types of initial conditions. In the present case the initial conditions play the role of a driving parameter which determines the formation and disappearance of such structures. 

Our consideration was limited to the case of regular perturbations in the initial distribution of velocity [see Eq. (\ref{4_nature})]. 
For this class of initial conditions we have shown how such space-time structures emerge and disappear during the evolution. However, the assumption about regular perturbations is not necessary; in the general case the perturbations may be stochastic, too. The distribution of density will look like random distribution, although this structure has been born by a deterministic process and our conclusions about the finite time of pseudo-chaotic behavior remain valid. Taking into account the high sensitivity of our pattern to perturbations in the initial conditions, one can come to the conclusion that there exists some interval in the system parameters [see Eqs. (\ref{5_n0a}), (\ref{8_n0a}) and (\ref{9_n0a})] in which it is possible to see the effect described in reality.

It should be noted that in the framework of the model the nonlinear effects (such as a density collapse or the temporal emergence and disappearance of wavelets) are, as a consequence, hence less pronounced since for its development there is not enough time available  within one period of a plasma oscillation. Nevertheless, we believe that our present approach can be generalized, for instance, to an inhomogeneous, many-component plasmas in fluid description \cite{ss,k02,sg}. Further generalizations of the model can be realized by some modifying the basis solution (\ref{cold_col})-(\ref{x_sh}). Such generalizations can also be used as a starting point for investigation of more complex and more realistic problems \cite{ir_1}-\cite{s96}.

In conclusion, we again would like to emphasize that the one-dimensional model, the transient nonlinear dynamics of which we have analyzed here rigorously on mathematical grounds, is still rather artificial because real systems are essentially higher dimensional and, in the absence of dissipation, kinetic. However, from the presented results we may catch a glimpse of the formation of time-dependent structures in higher dimensions governed by the hydrodynamic nonlinearity as a reference to a more realistic dynamical evolution, which generally involves particle trapping processes \cite{Sch12,Sch15}.

\end{document}